\documentclass{aip-cp}

\usepackage[numbers]{natbib}
\usepackage{rotating}
\usepackage{graphicx}

\begin{document}

\title{Neutrino emission, Equation of State and the role of strong gravity}

\author[aff1]{O. L. Caballero\corref{cor1}}

\affil[aff1]{Department of Physics, University of Guelph, Guelph, Ontario N1G 2W1, Canada.}
\corresp[cor1]{Corresponding author: ocaballe@uoguelph.ca}

\maketitle

\begin{abstract}
Neutron-star mergers are interesting for several reasons: they are proposed as the progenitors of short gamma-ray bursts, 
they have been speculated to be a site for the synthesis of heavy elements, and they emit gravitational 
waves possibly detectable at terrestrial facilities. The understanding of the merger process, from the pre-merger 
stage to the final compact object-accreting system involves detailed knowledge of numerical relativity and nuclear physics.
In particular, key ingredients for the evolution of the merger are neutrino physics and the matter equation of state. 
We present some aspects of neutrino emission from binary neutron star mergers showing the impact that the equation of state has 
on neutrinos and  discuss some spectral quantities relevant to their detection such as energies and luminosities far from the source.
\end{abstract}

% Head 1
\section{INTRODUCTION}
Together with gravity neutrinos drive the evolution of several stellar phenomena such as supernovae, binary mergers, and black hole accretion disks.
They are also a key ingredient in the synthesis of heavy elements \cite{Lattimer1976,Surmanrprocess}, the production of gamma ray bursts \cite{RuffertGRB-BH,Nakamura:2013bza}
and kilonova \cite{Piran:2014wpa,Metzger:2011bv}. Despite the numerous
efforts to fully understand these phenomena there are still puzzles to address. 
Unrevealing the missing pieces requires complex and computationally expensive simulations of the system under study that bring together details of nuclear physics and gravity.
Furthermore, due to the complexity of the problem, it is 
necessary to post-process the results to shed light on other derived aspects such as the element abundances and neutrino detection.

The strong gravitational field generated by a binary neutron star merger changes the spectral properties of the emerging neutrinos. This will affect any physical
quantity related to them. For example, the emitted neutrinos can interact with matter outflowing the merger. Via weak interactions neutrinos will change the neutron to proton ratio
(electron fraction) of the outflow
 setting the path that a nuclear reaction chain follows. However, strong gravity effects, such as redshifts and bending of neutrino trajectories, affect the neutrino fluxes 
 altering the electron fraction.
 As a consequence the final abundances differ from those obtained in the absence of the gravitational field \cite{GRCaballero}. 
On another hand, if the merger occurs in the Milky way or in the local galaxy group chances are that we will able to detect those neutrinos with current and future facilities
\cite{Nagatakicounts,McLaughlin07,Caballerosurface}. It is
therefore interesting to study the behavior of neutrinos emitted from binary systems.

In this proceedings we present post-processed neutrino emission results from
state of the art 3D fully relativistic binary neutron star simulations with magnetic fields, neutrino cooling and different equations of state. 
The original simulation results of the coalescence have been presented in reference \cite{Palenzuela:2015dqa}. There, the evolution of the matter in connection 
to the Equation of State (EoS) is discussed,
as well as the gravitational wave emission and neutrino detection rates. Here, we expand our discussion of neutrino emission focusing on the evolution of the neutrino surface and
its relation to the EoS.

\section{EQUATION OF STATE}
The EoS connects the macroscopic thermodynamical variables to the microphysics of the neutron star. Observables such as maximum masses and radii 
carry important information about the behavior of nuclear matter at high densities. Similarly, our theoretical models of nuclear interactions 
should correctly account for these observables. Among other macroscopic variables, the EoS predicts the temperature, density and matter composition.
The neutrino absorption and emission rates are correlated to these same quantities. A microphysical EoS is therefore fundamental to track the merger dynamics 
and the neutrino emission. We consider three different EoS (tables  publically available at www.stellarcollapse.org), NL3, DD2, SFHo, all of them 
in the frame of the statistical model of Hempel and Schaffner-Bielich \cite{Hempel:2009mc}. They differ in the relativistic mean field model used to describe the nuclear interaction. 
The NL3 EoS is based on the interaction model of reference \cite{Lalazissis:1996rd}, the DD2 on \cite{Typel:2009sy} and the SFHo on \cite{Steiner:2012rk}. These EoS predict different radii and maximum masses. The EoS 
is said to be \textit{soft} if results in a small radius for a given mass and is \textit{stiff} if for the same mass the predicted radius is larger. Our simulations
consider the coalescence of two neutron stars each with a mass of $1.35 M_\odot$. The SFHo predicts for that mass a neutron star radius of $\approx$ 12 km.
The DD2 predicts a close to $13$ km radius while for the NL3 the radius is $\approx 15$ km. 
In that order of ideas the SFHo is the softest of the EoS considered, the DD2 is an intermediate EoS and the NL3 is the stiffest one.  

\section{NEUTRINO EMISSION}

Post-merger neutrino luminosities can be as large as $10^{54}$ erg/s. This huge amount of neutrinos can reach detectors on Earth and tell us valuable information
about the merger and its evolution. At high matter densities ($\approx 10^{14}$ g/cm$^3$) neutrinos are trapped: the neutrino
mean free path $\lambda_\nu$ is shorter than the physical dimensions of the merger, i. e. matter is neutrino opaque. Once the density decreases the neutrino mean free path increases and neutrinos
escape carrying energy away and contributing to the cooling of the system. 
The point of last scattering, the place where neutrinos are free, is known as the neutrino surface. 
A common approach to determine this surface is to find the places where the optical depth $\tau =2/3$ (see e.g. \cite{Shapiro}). 
In terms of the opacity $\kappa_\nu$, and the mean free path $l_\nu$, the optical depth $\tau$ is given by
 \begin{equation}
\label{opticaldepth}
\tau_{\nu}=\int^{\infty}_{s_{\nu}}\kappa_\nu(s')ds'= \int^{\infty}_{s_{\nu}}\frac{1}{l_{\nu}(s')}ds',
\end{equation}
along the propagation direction $\hat{s}$. In this post-processing we chose to study the neutrino emission as seen from an observer located at infinity
in the $z$-axis. Then, the neutrino surfaces are defined by the two-dimensional hypersurfaces where the optical depth of
the merger is perpendicular to the equatorial plane. This turns the optical depth into

 \begin{equation}
\label{opticaldepthz}
\tau_{\nu}(x,y)=\int^{\infty}_{z_{\nu}}\frac{1}{l_{\nu}(x,y,z')}dz'.
\end{equation}
We integrate Equation \ref{opticaldepthz} changing the lower limit $z_\nu$ until the value $\tau=2/3$ is reached. The themodynamical conditions of the neutrino
surface are therefore the conditions of the matter at $z_\nu$. 

In Equation \ref{opticaldepthz} the mean free path depends of the cross sections $\sigma_k$, 
\begin{equation}
\label{meanfreepath}
l_\nu(x,y,z)=\frac{1}{\sum_kn_k\langle\sigma_k(E_\nu)\rangle}.
\end{equation}
 The sum above goes over all the relevant scattering and absorption processes $k$ that
 neutrinos undergo as they diffuse through matter, with $n_k$ the number density of the target. $\langle\sigma_k(E_\nu)\rangle$ is
the thermally averaged (``weighted'')
cross section,
\begin{equation}
\label{averagedcross}
\langle  \sigma_k(E_\nu)\rangle=\frac{\int ^\infty_0\sigma_k(E_\nu)\phi (E_\nu)dE_\nu} {\int^\infty_0\phi(E_\nu)dE_\nu},
\end{equation}
where $\phi(E_\nu)$ is the neutrinos Fermi-Dirac flux
\begin{equation}
\phi(E)=\frac{c}{2\pi^2(\hbar c)^3}E^2f_{FD}
\end{equation}
and $f_{FD}$ is the neutrino occupation number. We assume the temperature
is equal to the local temperature and zero chemical potential.
This procedure removes the energy dependence of the neutrino surface. Note, that
the scattering neutrino surfaces introduced above differ from the effective neutrino surfaces (see e.g. \cite{Perego:2014fma}).
The first ones correspond to the last scattering surfaces, where absorption and scattering neutrino processes
are treated equally while in the calculation of the effective neutrino
surfaces more emphasis is given to absorption processes.
The effective surfaces would be however diffuse in area but will give a better estimate
of the neutrino temperature, while the scattering surfaces will give a better estimate
of the effective area of decoupling.
As our calculation for average energies (explained below) is thermally weighted over the Fermi-Dirac distribution
we expect that they will be good estimates.

In this analysis we consider neutrino scattering from protons, neutrons and electrons,
as is a good approximation to assume that matter is dissociated at the typical values of $T$=10 MeV (and above)
found in simulations. 
In this way we have the charged current processes 
\begin{equation}
\nu_e +n\rightarrow p+e^-,
\label{nuabsorption}
\end{equation}
and
\begin{equation}
\label{anuabsorption}
\bar{\nu}_e +p \rightarrow e^+ + n.
\end{equation}
We also consider neutral current processes, elastic scattering from electrons and neutrino-antineutrino annihilation, which affect on equal footing all neutrino flavors. 
We find proton and neutron number densities assuming charge neutrality $Y_e=Y_p$,
and the electron number density assuming equilibrium of thermal electrons and positrons with radiation.
Details on the cross sections of the above reactions can be found in ref. \cite{Caballerosurface}.  As
the heavy lepton neutrinos, tau and muon, do not undergo the process of equations \ref{nuabsorption} and \ref{anuabsorption} they
are the first to decouple form matter. Also as the medium is neutron-rich, due to the original
neutron richness of the neutron stars, the last ones to decouple are electron neutrinos given the predominance of the reaction of equation \ref{nuabsorption}.

Using our results for the neutrino surfaces we can make estimates of the number
of neutrinos leaving the merger per second $dN/dt$ and average energies $\langle E_\nu\rangle$. For the first one we have
\begin{equation}
 \frac{dN}{dt}=\frac{c}{2\pi^2(\hbar c)^3}\int dA dE E^2f_{FD}=\int dA dE \phi(E),
\end{equation}
and for the energy rate (energy per sec)
\begin{equation}
\frac{dE}{dt}=\frac{c}{2\pi^2(\hbar c)^3}\int dA dE E^3f_{FD}=\int dA dE E \phi(E).
\end{equation}
In the equations above the integral over $dA$ corresponds to an integral over the neutrino surface.
Our estimate for the average neutrino energy is then given by
\begin{equation}
\langle E_\nu\rangle=\frac{dE/dt}{dN/dt}.
\end{equation}

In order to transform the average energy to another reference frame we make use of the fact that the quantity $I/c^3$, with $I$ the specific intensity is an invariant, 
\begin{equation}
 \frac{I}{E^3}=\frac{1}{c^2}\frac{dN}{d^3xd^3p}=\frac{f_{FD}}{h^3c^2}.
\end{equation}
Noting the quantities measured by an observer with ``tilde'', so the observed energy is $\tilde E$, and $d\tilde A$ is
the observed area differential, and without tilde analogous emitted quantities  $E$ and $dA$, then we have

\begin{equation}
  \frac{1}{c^2}\frac{dN}{d^3\tilde xd^3\tilde p}=\frac{1}{c^2}\frac{dN}{d^3xd^3p},
\end{equation}
which leads to

\begin{equation}
\frac{dN}{d\tilde t}=\int\frac{f_{FD}}{h^3c^2} \tilde E^2d\tilde E d \tilde A \tilde d\Omega.
\end{equation}

We can rewrite this in terms of the emitted quantities given that $\tilde E =\sqrt{g_{00}} E$, with $g_{00}$ the redshift (in the Schwarzchild metric for simplicity),
assuming that the emission is isotropical $d\tilde \Omega =4 \pi$, and that distances are stretched by a factor of $(1-r_s/r)^{-1/2}=g^{-1/2}_{00}$, 

\begin{equation}
\frac{dN}{d\tilde t}=\int g^{1/2}_{00} \phi(E)dEdA.
\end{equation}
We then get $\tilde E dN/d \tilde t=d\tilde E/d \tilde t$,
\begin{equation}
 \frac{d \tilde E}{d\tilde t}=\int g_{00} \phi(E)EdEdA,
\end{equation}
and the observed average neutrino energy is just the ratio of the above expressions, $\langle \tilde E_\nu\rangle=\frac{d\tilde E/d \tilde t}{dN/d \tilde t}$.

\subsection{Evolution of the neutrino surface: the DD2 EoS} 
Following the methodology of the previous section we find the neutrino surfaces for the three different neutrino flavors.
The electron antineutrino surface at $t=7.4$ ms after the merger for the DD2 EoS is presented in Figure \ref{eanuDD2}. The height represents the distance, measured from the equatorial plane, at which neutrinos decouple
from matter, whereas the color scale shows the matter temperature at these last points of neutrino scattering. With these temperatures we determine the 
average neutrino energies
as described above.
\begin{figure}[h]
  \centerline{\includegraphics[trim=2cm 2cm 0cm 0cm, clip=true, width=2. in, angle=-90]{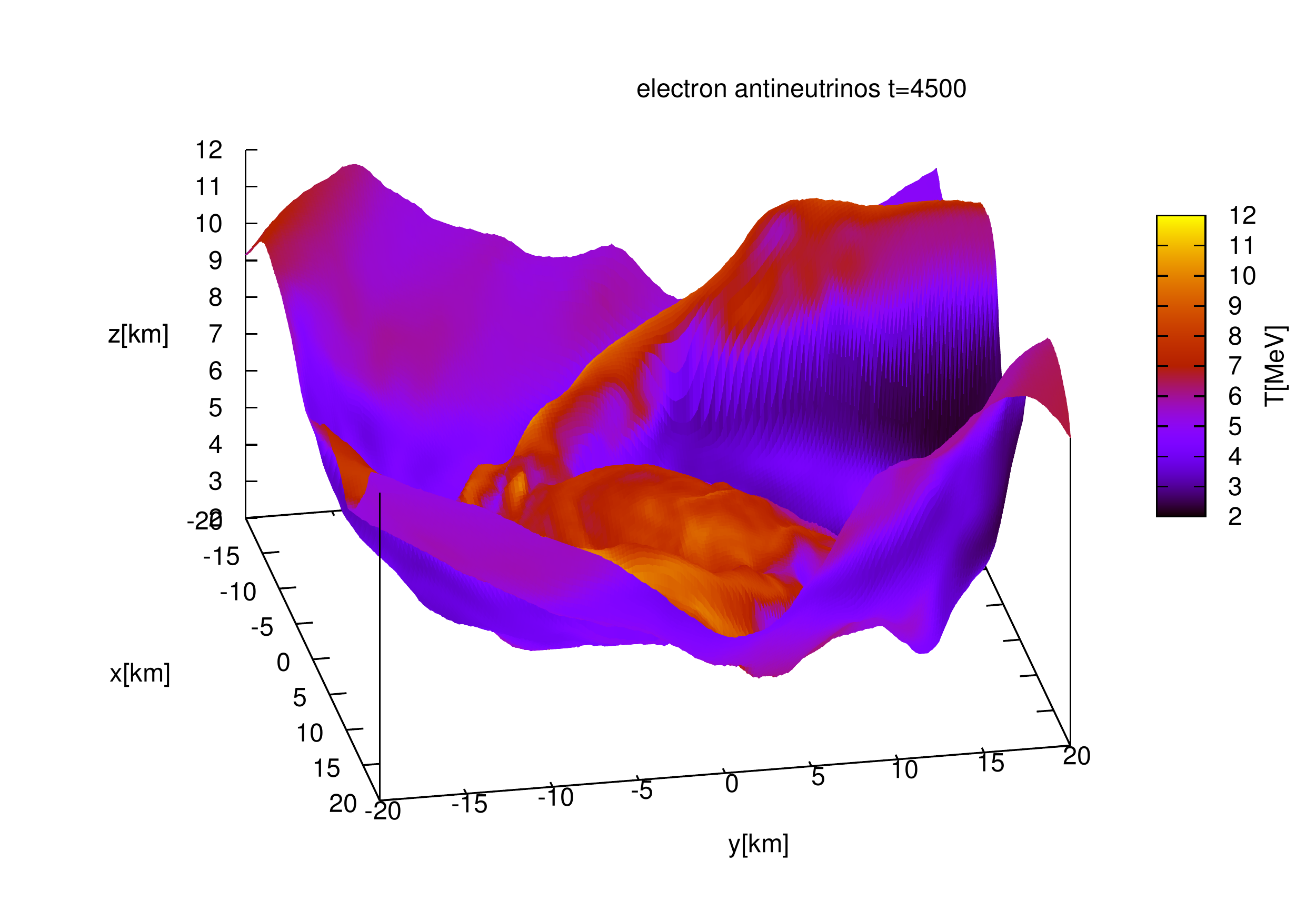}}
  \caption{The electron antineutrino surface at t=7.4 ms after the merger
for the DD2 EoS.}
\label{eanuDD2}
\end{figure}

Figure \ref{eanuDD2} shows the time evolution of the electron antineutrino surface for the DD2 EoS as seen from the $z$-axis. The time after the merger is 
as indicated in the figure. Although, electron antineutrinos achieve larger temperatures, around $18$
MeV, at earlier times near the center for t=2.5 ms after the merger, 
larger areas are warmed  up at latter times. The highest temperature
for electron antineutrino decreases $\sim$ 12 MeV at t= 7.4 ms, but matter is overall warmer leading to the larger
neutrino luminosities as reported in \cite{Palenzuela:2015dqa}. For even latter times, t=12.3 and 17.2 ms, the flux of neutrinos has contributed to the 
matter cooling and the neutrinos temperatures decrease. 
\begin{figure}[h]
  \centerline{\includegraphics[trim=0cm 1cm 0cm 0cm, clip=true, width=3.5 in, angle=-90]{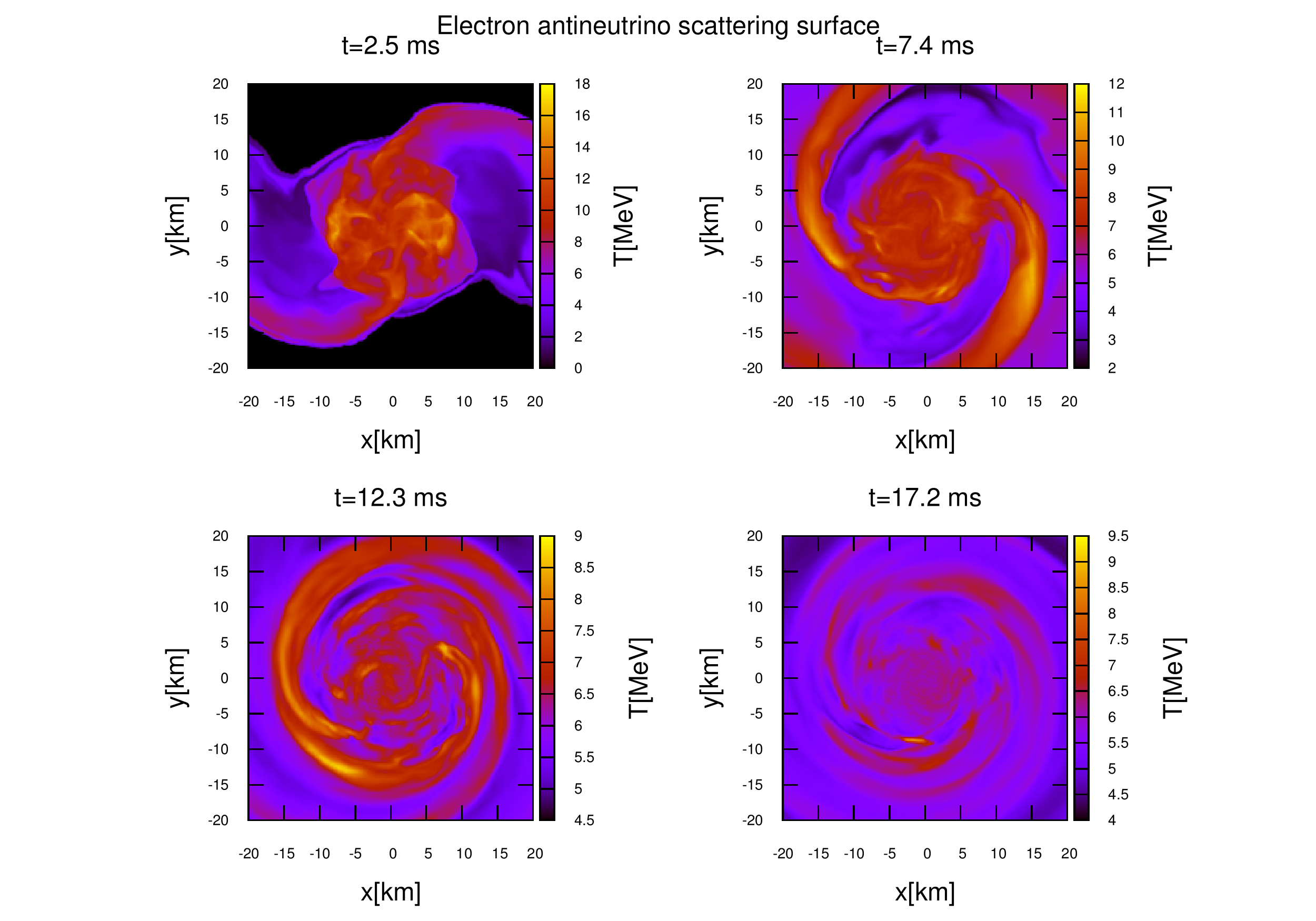}}
  \caption{A projection on the $z=0$ plane of the electron antineutrino surface at different times after the merger
for the DD2 EoS.}
\label{eanuDD2}
\end{figure}

Table \ref{energytable} shows the deduced average neutrino energies for electron
antineutrinos and electron neutrinos at the corresponding times after merger. In parenthesis are indicated the energies at the emission point while out of
the parenthesis are the observed quantities at infinity.

\begin{table}[h]
\caption{Observed(emitted) neutrino average energies for electron neutrinos and electron antineutrinos for the DD2 EoS.}
\tabcolsep7pt\begin{tabular}{lcccc}
\hline
 Time(ms)&$\langle E_{\bar\nu_e}\rangle$ &$\langle E_{\nu_e}\rangle$ \\
       & (MeV)&(MeV)\\
\hline

2.5 &18.3 (22.1)& 14.6 (17.4)\\
7.4& 13.2 (16.1) &10.2 (12.4)\\
12.3& 12.6 (15.3)&10.1 (12.3)\\
17.2& 10.7 (13) & 8.9 (10.8)\\
\hline
\end{tabular}
\label{energytable}
\end{table}

\subsection{Neutrino surfaces for different EoS}
\begin{figure}
 \centerline{\includegraphics[width=3.5 in,angle=-90]{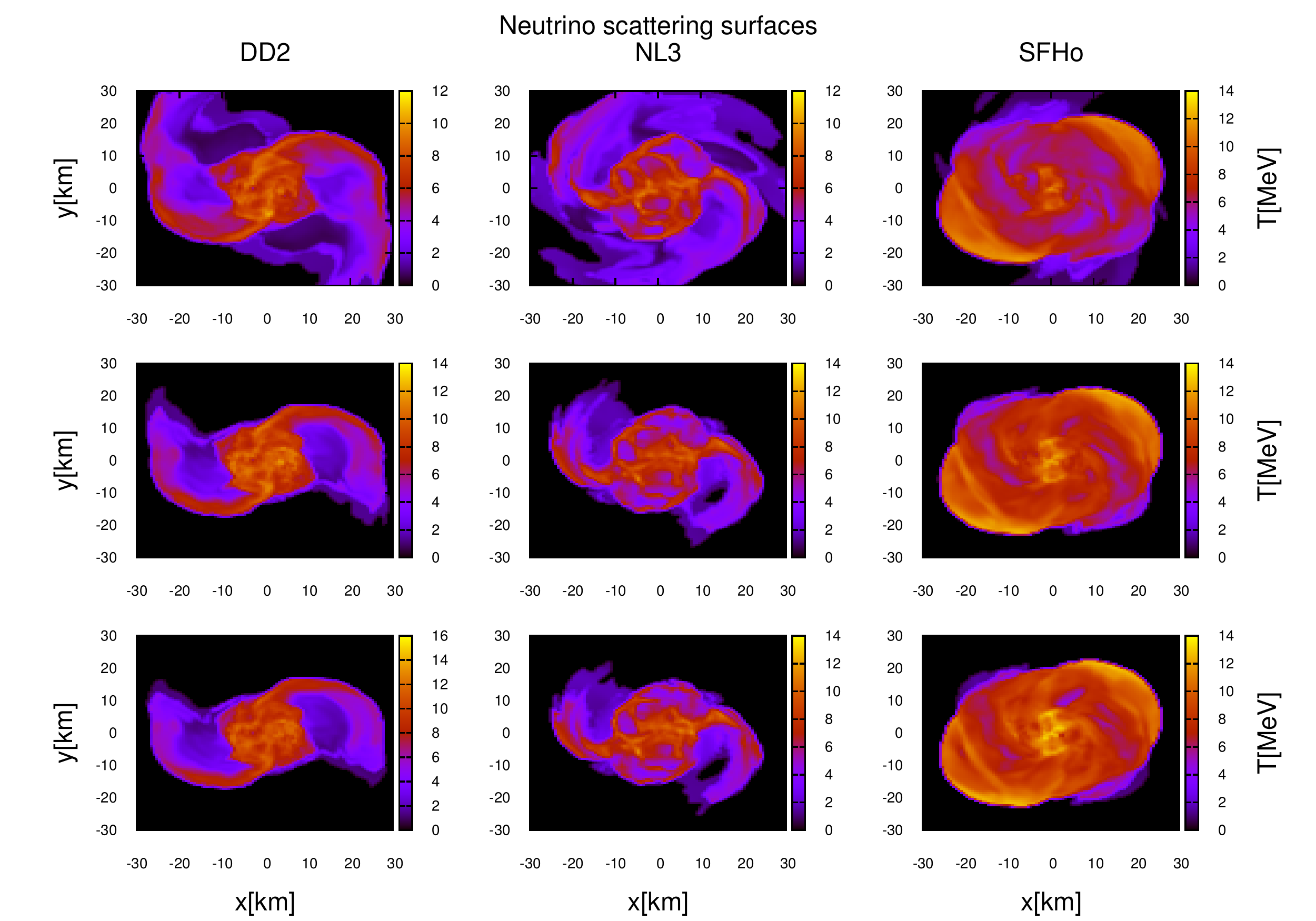}}
\caption{ Comparison of electron neutrino (top), electron antineutrino (middle), and tau/muon neutrino (bottom) surfaces 
at a time $t\sim$ 2.5 ms (see table \ref{EOSenergytable})
for different EoS.}
\label{nuEOS}
\end{figure}

Figure \ref{nuEOS} shows the neutrino surfaces for electron neutrino, electron antineutrino and tau/muon neutrino
for different EoS at similar times after the merger. The SFHo EoS results
in over all warmer larger surfaces which is a result of a more violent collision: the softer EoS allows smaller neutron star radii, therefore the merger
happens deeper in the gravitational potential. A fraction of this gravitational energy is converted to neutrino energy. In contrast the stiffest EoS NL3
predicts larger radii. This means the merger occur at earlier times and less deep in the potential. As a consequence the gravitational energy
available for conversion to neutrino energy is smaller compared to the SFHo case. The intermediate surfaces are the
DD2. We expect then that a neutrino signal will be more energetic for the SFHo, then DD2 and last for the NL3 EoS.
Table \ref{EOSenergytable} shows the average neutrino energies observed at infinity with our estimates. The values in the table are identical to those reported in table III of 
reference \cite{Palenzuela:2015dqa} and we write them here for completeness.

It is also expected (and can be seen in Figure 15 of Reference \cite{Palenzuela:2015dqa}) that the
neutrino surfaces for the other EoS will follow a similar time evolution of as the DD2 case discussed above. 
Before the merger the matter is cold and there is not significant neutrino emission. 
Soon after the merger matter is warmed up and neutrino production becomes important. Howerver, due to the high densities
neutrinos are trapped. As time passes the temperature increases and neutrinos leave the merger. Their luminosites
reach a maximum which depends on the EoS.
Between 4.5 ms and 15 ms for DD2, and between 3 and 9 ms for NL3, a quasi-stationary state is reached 
where the neutrino luminosities are slightly lower than the maximum value. However, for the SFHo EoS, this state is not reached as the
hyper massive neutron star, result of the merger, collapses to a black hole. Therefore, for the cases studied here neutrinos with larger energies
could be a signature of a softer EoS.

\begin{table}[h]
\caption{Observed at infinity neutrino average energies for electron neutrinos and electron antineutrinos for three different EoS.}
\tabcolsep7pt\begin{tabular}{lcccc}
\hline
 EoS & Time(ms)&$\langle E_{\bar\nu_e}\rangle$ &$\langle E_{\nu_e}\rangle$ \\
      & & (MeV)&(MeV)\\
\hline

DD2&3.0 & 18.3  &14.6\\
DD2&7.9 & 13.2 &10.2\\
NL3&2.5 &18.5 &15.2 \\
NL3&8.4 & 13.4 &9.8\\
SFHo& 3.2&24.6&23.5\\

\hline
\end{tabular}

\label{EOSenergytable}
\end{table}

\section{CONCLUSIONS}

Binary neutron star mergers represent a wonderful laboratory to contrast our theoretical predictions with observations. From one side the occurrence of an event
will produce a gravitational wave signal sufficiently strong to be detected at existent facilities such as Advanced LIGO. On the other hand, we will be also able to detect 
neutrinos emitted during the merger with current and future water Cherenkov detectors (such as Super Kamiokande). Furthermore, we expect that from these mentioned
signals we will gain information from the EoS and consequently about the nuclear force. In this proceedings we have discussed the influence that the EoS has on the
neutrino emission from binary neutron star mergers.

We have followed the evolution of the neutrino surface for the DD2 EoS, and found that for earlier times the neutrino production is low due to the higher densities of the system. 
At later times (about 2.5 ms after merger) the  neutrino production peaks as the matter distribution is more even and the temperatures increase. At this stage
the neutrino surface temperatures are high and neutrinos have an average energy at infinity around 18 MeV. Finally, the emission decreases as the system losses
energy via neutrinos. This quasi-steady state last for some milliseconds.  
We showed that the stiffness of the EoS changes the neutrino energies. A soft EoS like the SFHo will produce hotter and higher energy neutrinos compared to
the stiffer NL3 EoS. For all cases the neutrino emission will last tens of milliseconds. This together with the higher neutrino energies of the merger will allow us to 
distinguish this signal from a Supernovae one which is estimated to be around 10 seconds.

Fully relativistic hydrodynamical simulations with realistic EoS represent a valuable method to understand the evolution of neutron star
mergers. Together with an accurate description of neutrino fluxes and nuclear reaction 
networks that determine the elements produced during these events will shed light on our understanding of the production of heavy elements. 
Future efforts will be focused on this problem.

\section{ACKNOWLEDGMENTS}
We thank our collaborators L. Lehner, C. Palenzuela, S. L. Liebling, D. Neilsen, E. O'Connor, and M. Anderson.


\nocite{*}


\begin{thebibliography}{99}

\bibitem{Lattimer1976} J. M. Lattimer and D. N. Schramm, ApJ, {\bf 210}(1976) 549.
\bibitem{Surmanrprocess} R. Surman and G. C. McLaughlin, ApJ, {\bf 679} (2008) L117.

\bibitem{RuffertGRB-BH} M. Ruffert, H.-Th. Janka, Astron. Astrophys.,  {\bf 344} (1999) 573.

  \bibitem{Nakamura:2013bza}
  K.~Nakamura, S.~Harikae, T.~Kajino and G.~J.~Mathews,
  %``r-process nucleosynthesis in the neutrino-heated relativistic collapsar jet model for gamma-ray bursts,''
  PoS NICXII {\bf } (2012) 216.

%\cite{Piran:2014wpa}
\bibitem{Piran:2014wpa} 
  T.~Piran, O.~Korobkin and S.~Rosswog,
  %``Implications of GRB 130603B and its macronova for r-process nucleosynthesis,''
  arXiv:1401.2166 [astro-ph.HE].
  
%\cite{Metzger:2011bv}
\bibitem{Metzger:2011bv} 
  B.~D.~Metzger and E.~Berger,
  %``What is the Most Promising Electromagnetic Counterpart of a Neutron Star Binary Merger?,''
  Astrophys.\ J.\  {\bf 746}, 48 (2012)
  doi:10.1088/0004-637X/746/1/48
  [arXiv:1108.6056 [astro-ph.HE]].  
  
\bibitem{GRCaballero} O. L. Caballero, G. C. McLaughlin and R. Surman ApJ, {\bf 745}, 170 (2012)
\bibitem{Nagatakicounts} Shigehiro Nagataki and Kazunori Kohri, Prog. Theor. Phys., {\bf 108} (2002)789.
\bibitem{McLaughlin07} G. C. McLaughlin and R. Surman, PRD, {\bf 75} (2007) 023005.
\bibitem[Caballero et al.(2009)]{Caballerosurface} O. L. Caballero, G. C. McLaughlin, \& R. Surman, Phys.\ Rev.\ D, {\bf 80} 123004 (2009)

%\cite{Palenzuela:2015dqa}
\bibitem{Palenzuela:2015dqa} 
  C.~Palenzuela, S.~L.~Liebling, D.~Neilsen, L.~Lehner, O.~L.~Caballero, E.~O'Connor and M.~Anderson,
  %``Effects of the microphysical Equation of State in the mergers of magnetized Neutron Stars With Neutrino Cooling,''
  Phys.\ Rev.\ D {\bf 92}, no. 4, 044045 (2015)
  doi:10.1103/PhysRevD.92.044045
  [arXiv:1505.01607 [gr-qc]].

%\cite{Hempel:2009mc}
\bibitem{Hempel:2009mc} 
  M.~Hempel and J.~Schaffner-Bielich,
  %``Statistical Model for a Complete Supernova Equation of State,''
  Nucl.\ Phys.\ A {\bf 837}, 210 (2010)
  doi:10.1016/j.nuclphysa.2010.02.010
  [arXiv:0911.4073 [nucl-th]].
%\cite{Hempel:2011mk}

%\cite{Lalazissis:1996rd}
\bibitem{Lalazissis:1996rd} 
  G.~A.~Lalazissis, J.~Konig and P.~Ring,
  %``A New parametrization for the Lagrangian density of relativistic mean field theory,''
  Phys.\ Rev.\ C {\bf 55}, 540 (1997)
  doi:10.1103/PhysRevC.55.540
  [nucl-th/9607039].
  
%\cite{Typel:2009sy}
\bibitem{Typel:2009sy} 
  S.~Typel, G.~Ropke, T.~Klahn, D.~Blaschke and H.~H.~Wolter,
  %``Composition and thermodynamics of nuclear matter with light clusters,''
  Phys.\ Rev.\ C {\bf 81}, 015803 (2010)
  doi:10.1103/PhysRevC.81.015803
  [arXiv:0908.2344 [nucl-th]].
  
  
%\cite{Steiner:2012rk}
\bibitem{Steiner:2012rk} 
  A.~W.~Steiner, M.~Hempel and T.~Fischer,
  %``Core-collapse supernova equations of state based on neutron star observations,''
  Astrophys.\ J.\  {\bf 774}, 17 (2013)
  doi:10.1088/0004-637X/774/1/17
  [arXiv:1207.2184 [astro-ph.SR]].
  
\bibitem{Shapiro} S. L. Shapiro and S. A Teukolsky, {\it Black Holes, White Dwarfs and Neutron Stars}, Wiley, New York, 1983.  
\bibitem{Perego:2014fma} 
  A.~Perego, S.~Rosswog, R.~M.~Cabezón, O.~Korobkin, R.~Käppeli, A.~Arcones and M.~Liebendörfer,
  %``Neutrino-driven winds from neutron star merger remnants,''
  Mon.\ Not.\ Roy.\ Astron.\ Soc.\  {\bf 443}, no. 4, 3134 (2014)
  [arXiv:1405.6730 [astro-ph.HE]].
\end{thebibliography}
\end{document}